\documentclass{JHEP}
\usepackage{epsfig}
\usepackage{amssymb}
\usepackage[active]{srcltx}

\newcommand{\bea}{\begin{eqnarray}}
\newcommand{\eea}{\end{eqnarray}}
\newcommand{\bean}{\begin{eqnarray*}}
\newcommand{\eean}{\end{eqnarray*}}

\def\O #1{\overline{#1}}

\def\ba{\begin{array}}
\def\ea{\end{array}}
\def\beq{\begin{equation}}
\def\eeq{\end{equation}}

\def\Tr{\mathop{\rm Tr}}
\def\det{\mathop{\rm det}}

\def\a{{\alpha}}
\def\b{{\beta}}
\def\d{{\rm d}}

\def\th{{\theta}}
\def\bth{{\overline{\theta}}}

\def\half{{1 \over 2}}
\def\Th{{\Theta}}

\preprint{SNUST-030701\\ {\tt hep-th/0307091}}
\title{Non(anti)commutative Superspace, \\
UV/IR Mixing \& Open Wilson Lines}
\author{Ruth Britto ${}^a$, Bo Feng ${}^a$, Soo-Jong Rey
${}^{a,b}$\\
~~~~\\
${}^a$ Institute for Advanced Study \\
Einstein Drive, Princeton NJ 08540 \\
~~~~~~~~~~~~~~~~\\
${}^b$ School of Physics \& BK-21 Physics Division \\
Seoul National University, Seoul 151-747 KOREA  \\
~~~~\\
\email{britto, fengb@ias.edu, sjrey@gravity.snu.ac.kr} }
\abstract{We study quantum aspects of field theories defined on
${\cal N}={1 \over 2}$ superspace, where both bosonic and
fermionic coordinates are made non(anti)commutative.  We compute
the one-loop effective superpotential, and we find that planar and
nonplanar contributions exhibit markedly different behavior.
Planar diagrams yield an effective superpotential proportional to
$N_c (\Phi \log \Phi)_\star$. For nonplanar diagrams, we show that
ultraviolet-infrared mixing takes place and explain why some
nonplanar diagrams are ultraviolet-divergent when bosonic
noncommutativity is turned off. Each nonplanar diagram is not
expressible as a star product of background fields, but, once
resummed appropriately, they are expressed as a star product
involving open Wilson lines in superspace. The open Wilson lines are
responsible for ultraviolet-infrared mixing. We comment on an intriguing
relation of our result to the Dijkgraaf-Vafa correspondence
between gauge theories and matrix models. }
\begin{document}
\newpage
%%%%%%%%%%%%%%%%%%%%%%%%%%%%%
\section{Introduction}
%%%%%%%%%%%%%%%%%%%%%%%%%%%%%
D-brane dynamics in the background of a closed string $p$-form gauge
field has been a source of surprises. When a flat potential of the
Kalb-Ramond field $B_{NS}$ is turned on, it was discovered that
open string dynamics perceives noncommutative spacetime
\cite{connesdouglasschwartz}, whose coordinates obey the
Heisenberg algebra,
\bea \left[ y^m, y^n \right] = i \Theta^{mn}. \label{xnon} \eea
Moreover, in the Seiberg-Witten scaling limit
\cite{seibergwitten}, excitations of all closed string modes and
massive open string modes are decoupled from the low-energy
dynamics on the D-brane. As a result, there emerges on the D-brane
worldvolume a new kind of theories, referred to as noncommutative
field theories. These theories are now known to exhibit many
surprising features, such as ultraviolet(UV)-infrared(IR) mixing
\cite{uvir}, nonlocal open Wilson lines as physical observables
\cite{owl} and as a sort of master fields. With these features,
noncommutative field theories depart from the ordinary field
theories but behave more like fundamental string theories.

Given that noncommutative space emerged from (super)string
theories, can non(anti)commutative superspace emerge from
superstring theories? Recently, in the context of the Dijkgraaf-Vafa
correspondence \cite{dv} relating four-dimensional ${\cal N}=1$
supersymmetric gauge theories and zero-dimensional matrix models, it
was suggested that non(anti)commutative superspace indeed emerges
on D-brane worldvolume if one turns on self-dual graviphoton field
strength in four dimensions \cite{oogurivafa, seiberg,
berkovitsseiberg} or, more generally, Ramond-Ramond 2-form field
strength in ten dimensions \cite{stonybrook}. The Grassmann-odd
coordinates $\theta^\alpha (\alpha=1,2)$ are now
non(anti)commuting \cite{non-theta,kawai,seiberg} and obey a
Clifford algebra:
\bea \label{theta-non} \{ \theta^\alpha, \theta^\beta \}
= C^{\alpha\beta}, \eea

The development prompts the  study of quantum field theories defined on
non(anti)commutative superspace.
In
\cite{seiberg}, it was pointed out that these theories constitute
non(anti)commutative supersymmetric field theories, in which
the ordinary product is replaced by a $\star$-product:
\bea \star = \exp \left( -{i \over 2} \Th^{mn}
\overleftarrow{\partial \over \partial
y^m}\overrightarrow{\partial \over \partial y^n} -{1 \over 2}
C^{\a\b} \overleftarrow{\partial \over
\partial \th^\a}\overrightarrow{\partial \over
\partial \th^\b} \right). \label{starproduct} \eea
It was also pointed out that, with non(anti)commutativity turned on,
the ${\cal N}=1$ supersymmetry in four-dimensional Euclidean space
is broken to ${\cal N}=1/2$ supersymmetry, but nevertheless 
preserves the 
antichiral ring structure. In previous work
\cite{bfr}, we studied quantum aspects of non(anti)commutative
field theories (see also \cite{ty}), and established the ${\cal N}={1/
2}$ (non)renormalization theorem. According to the theorem, the
antiholomorphic superpotential ($\O {\rm F}$-term) is not renormalized,
while the holomorphic superpotential (F-term) is subject to
renormalization and is combined to the K\"ahler potential
(D-term). Nevertheless, the energy density of supersymmetric vacua
still vanishes to all orders in perturbation theory.

Being subject to renormalization, the superpotential (F-term) is
radiatively corrected by terms like $\Phi Q^2 \Phi$. As it stands,
such a term is not expressible in terms of the $\star$-product
Eq.(\ref{starproduct}), as the generic $\star$-product among
superfields produces only {\sl even} powers of $Q^2$. One might
regard it (as in \cite{ty}) as indicating that the $\star$-product
Eq.(\ref{starproduct}) one started with is no longer valid
quantum-mechanically, and hence needs to be modified to some sort of
generalized $\star$-product in the effective superpotential. Quite to the
contrary, in this work, we show that the $\star$-product
Eq.(\ref{starproduct}) performs just as well at the quantum level; rather,
the operators need to be reorganized to a set of nonlocal objects
called open Wilson lines. Specifically, in the effective
superpotential, nonplanar diagrams produce local operators of
increasing powers of the $\Phi$ field and $Q^2$ (as well as $\Box$'s)
acting on them, of which the aforementioned $\Phi Q^2 \Phi$ is one
of the lowest dimension operators. Each local operator is indeed
not expressible in terms of the $\star$-product
Eq.(\ref{starproduct}), but a suitable resummation of a set of
local operators is. We thus resolve the conundrum by demonstrating
that, by resumming individual terms, the effective superpotential
may be organized as a generating functional of the open Wilson
lines. Importantly, in defining the open Wilson line, no
modification to the $\star$-product Eq.(\ref{starproduct}) is
necessary.

Essentially the same conundrum  arose in the context of
noncommutative field theory \cite{star-x}, where, initially, a
class of generalized $\star$-products was considered inevitable
for expressing quantum effects. It was then found in
\cite{wilsonline} that the quantum effect is not to modify the
definition of the $\star$-product, but to reorganize the local
operators into  open Wilson lines. It is now well understood
that the open Wilson line, and the intuitive picture of it as an analog
of electric dipole in a magnetic field, are the fundamental reasons
that underlie the UV-IR mixing in ordinary noncommutative field
theories \cite{review}.

As said, the crux of the present work is to demonstrate that quantum
effect in non(anti)commutative field theories is not to modify the
$\star$-product, but to reorganize local operators into open Wilson
lines. For simplicity, in this work, we shall consider the ${\cal
N}={1 \over 2}$ Wess-Zumino model whose holomorphic action is
given by
\bea S = \int \d^4 y \d^2 \th \, \left[ {1 \over 2} \Phi \star
\left(  ({1 \over 4\lambda } Q^2) - {\Box \over \O m} + m \right)
\star \Phi + {g \over 3} \Phi\star\Phi\star\Phi + \cdots \right],
\nonumber \eea
in a suitable kinematical limit. We find, 
significantly, that the open Wilson lines extend over the
superspace.  So, for example, the open Wilson line carrying
supermomentum $(k, \kappa)$ takes the form\footnote{In this
expression, $h =(g / 2)^2 (\O m / m) |\Theta k|^2$ is a
kinematic factor defining a sort of the metric of the
$t$-parameter space. See below.}
\bea \label{susy-wilsonline}
 {\cal W}_{k, \kappa}[\Phi]= {\cal P}_\sigma
\int  \d^4 y\d^2 \theta \, \exp_\star \left( - \int_0^1 \sqrt{h}
\,\d \sigma \, \Phi(y+\Theta k \, \sigma, \, \theta- C \kappa \,
\sigma) \right) \star e^{-i k y - \kappa \theta}. \eea
Notice that the open Wilson line extends in $(y,\theta)$
superspace over the interval $(\Theta k,\, C\kappa)$, and hence
encodes the ${\cal N}={1 \over 2}$ version of the aforementioned {\it
dipole relations}:
\bea \label{dipole} \Delta y^m =\Theta^{mn} k_n \qquad {\rm and}
\qquad \Delta \theta^\alpha = C^{\alpha\beta} \kappa_\beta. \eea
Noting that $(k, \kappa)$ refers to the total momentum of the open
Wilson line, it is evident from these relations that there emerges
a UV-IR relation in {\sl superspace}.

This paper is organized as follows.  In section 2, we formulate
the holomorphic Wess-Zumino model Eq.(\ref{holoWZ}) by integrating
out the $\O\Phi$ field. We set up the one-loop background field method
and derive Feynman rules thereof. In section 3, we compute the
effective superpotential using the background field method. In
section 4, we explain our result and discuss in detail various
limits of interest, where the non(anti)commutativity is turned
off. In section 5, we resum planar diagrams and show that they
give rise to a contribution proportional to $N_c (\Phi \log
\Phi)_\star$. Appendix A contains notation, conventions, and some
details of Fourier transformation in ${\cal N}={1 \over 2}$
superspace. Appendix B contains the detailed derivation of various
results sketched in section 3.
%%%%%%%%%%%%
\section{ ${\cal N}= 1/2$ SUSY Wess-Zumino model}
%%%%%%%%%%%%
We begin with recollecting that the non(anti)commutative
Wess-Zumino model is defined as the ordinary Wess-Zumino model in
deformed superspace, whose coordinates obey Heisenberg/Clifford
algebras, Eqs.(\ref{xnon}, \ref{theta-non}):
\bea S =\int \d^4 y \left[ \int \d^2 \theta \d^2 \O\theta \, \Phi
\star \O\Phi + \int \d^2 \theta \Big( {m\over 2} \Phi \star \Phi
+{g\over 3} \Phi \star \Phi \star \Phi \Big) + \int \d^2 \O\theta
\Big( {\O m\over 2} \O\Phi \star \O\Phi +{\O g\over 3} \O\Phi
\star \O\Phi \star \O\Phi \Big) \right], \nonumber \eea
where the $\star$-product is as defined in Eq.(\ref{starproduct}).
We have proven in \cite{bfr} that the antiholomorphic
superpotential $\O W_\star (\O \Phi)$ is not renormalized to all
orders in perturbation theory. We will therefore study the effective
holomorphic superpotential $W^{\rm eff}_\star (\Phi)$ that is
generated by quantum effects. The situation is similar to
the diagrammatic derivation \cite{grisaru} of the Dijkgraaf-Vafa
correspondence between gauge theory and matrix model: both cases
concern computation of quantities governed and generated by the
holomorphic superpotential. As in \cite{grisaru}, to simplify the
computation, we will set $\O g$ (as well as couplings of
all higher monomials) to zero. We shall comment below on the consequences when
these antiholomorphic nonlinear couplings are non-vanishing.
Integrating out the $\O \Phi$ field amounts to doing a Gaussian integral (the
manipulation is standard and elementary, as the D-term is just a
nonchiral coupling to the `external source' $\Phi$), and results in a
term proportional to $\Phi \Box \Phi$. We will also add the ${\cal
N}=1/2$ kinetic multiplet term, $\Phi Q^2 \Phi$. In \cite{bfr}, we
demonstrated that this term is generated at one loop. In
fact, these two are the most general kinetic terms preserving
${\cal N} = \half$ supersymmetry. Putting it all together, the action of
holomorphic Wess-Zumino model may be written as
\bea S=\int \d^4 y \d^2 \theta  \left[ {1\over 2} \Phi(y,\theta)
\left({1 \over 4\lambda} Q^2 -{1 \over \O m}\Box + m \right)
\Phi(y,\theta)+{g\over 3} \Phi(y,\theta) \star
 \Phi(y,\theta) \star\Phi(y,\theta) \right]. \label{holoWZ}
\eea
Here, we have introduced a dimensionless coupling constant
$\lambda$ to the kinetic multiplet term, and we will treat it as a
variable parameter along with the others, $m, \O m$ and $g$.

In computing $N$-point one-particle-irreducible Green functions, as
in noncommutative field theories, we find it convenient to work in
momentum superspace and write the $\star$-product as a
momentum-dependent phase factor. The $\Phi$-field propagator is
given by
\bea \langle
\widetilde{\Phi}(k,\kappa)\widetilde{\Phi}(k',\kappa')\rangle & =
&  \Delta(k, \kappa) (2 \pi)^4 \delta^4(k+k') ({i \over
2})^2\delta^2(\kappa+\kappa') \nonumber \eea
where
\bean \Delta(k, \kappa) = {1 \over {1 \over \O m} k^2 +{1 \over 4
\lambda} \kappa^2 +m }. \eean
Denote\footnote{Our notation and conventions are explained in
detail in Appendix A.} superspace coordinates as $Y=(y,\theta)$
and corresponding momenta as $K=(ik,\kappa)$. Then the action
Eq.(\ref{holoWZ}) can be written in momentum space as
\bea S & = & {1\over 2} \int \prod_{i=1,2}\d^6 K_i
\widetilde{\Phi}(K_i) \, \Delta^{-1} (K_1) \, \delta^6(K_1+K_2)\nonumber \\
&+&{ g\over 3} \int \prod_{i=1,2,3} \d^6 K_i \widetilde{\Phi}(K_i)
 \,\, e^{\sum_{i<j} {i \over 2}  K_i \wedge K_j} \,
\delta^6(K_1+K_2+K_3)\label{mmholoWZ}. \eea

Two remarks are in order. \hfill\break
$\bullet$ As in \cite{grisaru}, we have set $\O g$ to zero. This
is only a mild convenience and not a severe restriction.
Indeed, suppose we leave $\O g$ finite and integrate out $\O
\Phi$. We then get an infinite series of additional terms
containing three or higher powers of $\Phi$. They are generated by
connected tree diagrams of $\O \Phi$'s, with propagator $1/\O m$ and
interaction vertex $\O g$, upon converting external $\O \Phi$
fields to $\Phi$ fields via the D-term $[\Phi \O
\Phi]_{\th\th\bth\bth}$. In doing so, we find that all these terms
involve a Laplacian $\Box_y$ and multiples of $Q^2$'s acting on
$\Phi$ fields. Thus, with $\O g$ nonzero, the holomorphic
superpotential $W_\star(\Phi)$ is not modified, and only
interaction terms involving derivatives are generated. As they
will either produce extra contact terms or higher-derivative
corrections to the open Wilson line, and do not entail any new
physics to the results we derive, we do not consider
these terms further. \hfill\break
$\bullet$ It is straightforward to extend the superfield $\Phi$
to a matrix-valued one and couple it to an external gauge field. In
particular, if $\Phi$ belongs to the adjoint representation of
the  gauge group $U(N)$ (or the bifundamental representation of the quiver gauge
group), the holomorphic Wess-Zumino model is equivalent via
Eguchi-Kawai reduction to a super-matrix model.\footnote{ This is in
spirit related to the observation made in \cite{kawai}.}  In this
case, the superspace integrals in the action and open Wilson lines
include a trace over the color indices of the $\Phi$-field.

Computation of the effective action at one loop is best facilitated 
by the background field expansion. Split
the superfield $\Phi$ into classical background and quantum
fluctuation:
\bea \Phi(Y) = \Phi_0(Y)+\varphi(Y) . \nonumber \eea
With fully symmetrized labelling of the momentum, the interaction
term is given by (keeping only the quadratic term of $\phi$)
\bea & & {g } \int \d^6 K_1 \cdots \d^6 K_3 \delta^6(K_1
+K_2+ K_3) \left[
%e^{-{1\over 2} \kappa_2\wedge \kappa_3 -{i\over 2} k_2\wedge k_3}
%+ e^{+{1\over 2} \kappa_2\wedge \kappa_3+{i\over 2} k_2\wedge k_3}
U+T\right] \varphi(K_1) \Phi_0(K_2) \varphi(K_3),
\label{interaction} \eea
where $U, T$ are phase factors:
\bea U(K_1, K_2) = \exp\left(- {i \over 2} k_1 \wedge k_2 - {1
\over 2} \kappa_1 \wedge \kappa_2 \right) \qquad T(K_1, K_2) =
\exp \left(+ {i \over 2} k_1 \wedge k_2 + {1 \over 2} \kappa_1
\wedge \kappa_2 \right). \,\, \label{u,t}\eea
The two phase factors differ by a relative sign in the exponent,
and we will call them the `untwisted' and the `twisted' vertex,
respectively. They are Hermitian conjugates. Notice that
the relative signs of bosonic and fermionic momentum phase factors
in $U, T$ are {\it correlated}, so they are untwisted or twisted
simultaneously. This property will be the crucial ingredient in
our computations.

%%%%%%%%%%%%%%%%%%%%%%%%%%
\section{The One-Loop Effective Superpotential}
%%%%%%%%%%%%%%%%%%%%%%%%%%
The action Eq.(\ref{holoWZ}), or equivalently, Eq.(\ref{mmholoWZ}), is
our starting point for the one-loop computation of the effective
superpotential in the $\Phi$-field background. As mentioned, we
will utilize the background field expansion around $\Phi =
\Phi_0(Y)$.
\EPSFIGURE[ht]{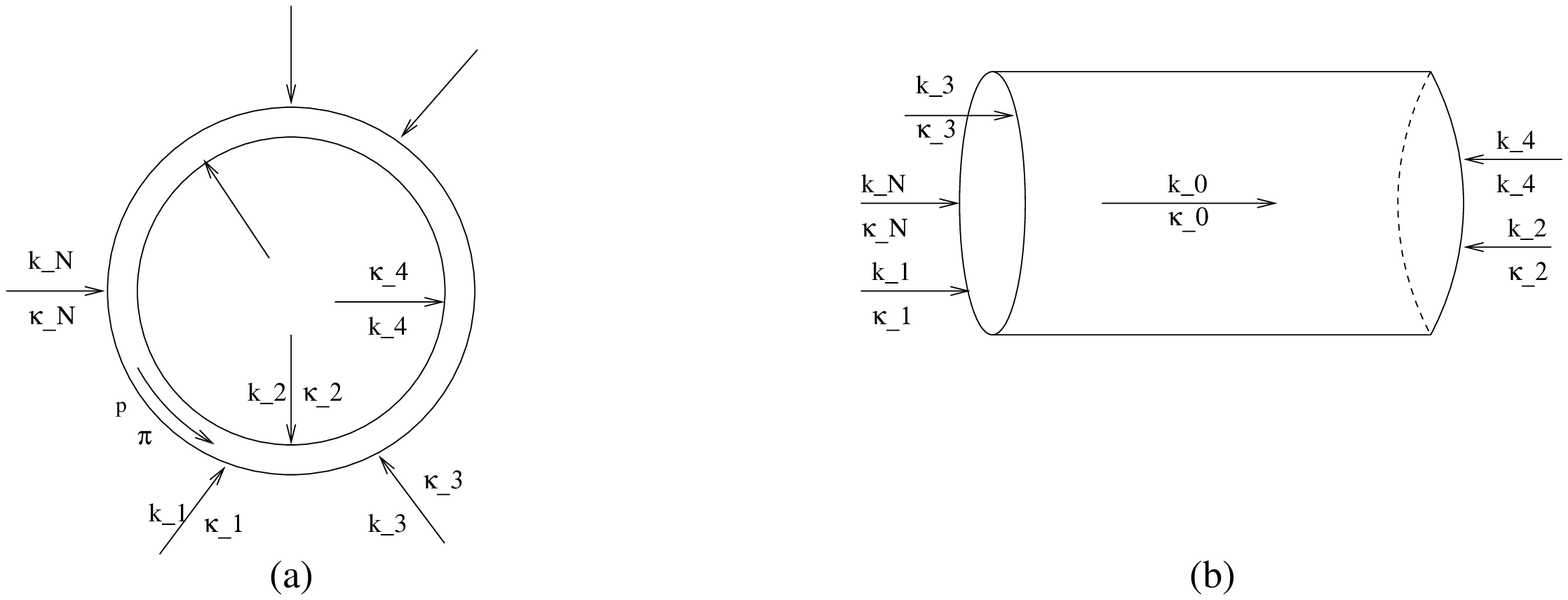,width=14cm} {(a) The one loop N-point function.  Double-line notation is featured in the loop to distinguish twisted and untwisted vertices. (b) Stretching the cylinder illustrates our eventual understanding in terms of open Wilson lines for untwisted and twisted vertices separately which hints at underlying open-closed duality. \label{f:oneloop} }
The one-loop computation proceeds as follows. Integrating out the
fluctuation field $\varphi$ yields schematically
\bea {\rm Tr} \int \d^6 K \log \left[ \Delta^{-1}(K) + \star\, g
\widetilde{\Phi}_0 \star \right]. \nonumber \eea
Expanding it in powers of the background field $\Phi_0$ gives
rise to insertion of the interaction vertices to the one-loop
vacuum diagram. Denote the super-momentum circulating the loop as
$P = (ip, \pi)$. The $N$-point function follows from the expansion
as
\bea G_{N} & = & (-{1\over N!}) (-{g})^N
\int \d^6K_1 \cdots \d^6 K_N  \, \widetilde{\Phi}_0(K_1) \cdots
\widetilde{\Phi}_0(K_N) \, \delta^6(K_1 + \cdots + K_N)\nonumber \\
&\times& \int \d^6 P  \prod_{i=1}^N \Delta \left(p + \sum_{j=1}^i
k_j, \pi + \sum_{j=1}^i \kappa_j \right)
\nonumber \\
&\times& \prod_{i=1}^N \exp{ \left[ {\epsilon_i \over 2}
\left(\kappa_i \wedge(\pi+\sum_{j=1}^{i} \kappa_j) + i k_i
\wedge(p +\sum_{j=1}^{i} k_j)\right) \right]}, \label{npt}
\eea
where $\epsilon=\pm 1$ is the relative sign for the un/twisted
vertex. The first line contains vertex symmetry factors, coupling
parameter, background fields, and overall super-momentum
conservation. The second line is the loop momentum integration
times $N$ propagators connecting adjacent background field vertex
pairs, and the third line is the phase factor $U$ or $T$
originating from the $\star$-product.

We shall use the double-line notation for the propagators and the
interaction vertices to distinguish the untwisted/twisted
vertices. See Fig.1(a). In this notation, the untwisted/twisted
vertex injects momentum to the inner/outer boundary. Evidently,
there are $2^N$ combinatorial possibilities, each vertex being
either untwisted or twisted. For each diagram contributing to the
N-point function, define the net momentum flow to the inner boundary as
\bean k_0 \equiv \sum_{i \in \{U\}} k_i, \qquad \kappa_0 \equiv
\sum_{i \in \{U\}} \kappa_i, \eean
where $i \in \{U\}$ denotes those vertices $i$ chosen to be untwisted. 
Overall momentum conservation implies that they are the same
as {\sl minus} the net momentum flow to the outer boundary:
\bean k_0 = -\sum_{i \in \{T\}} k_i, \qquad \kappa_0 = -\sum_{i
\in \{T\}} \kappa_i. \eean

Our next task is to perform the loop momentum integrals and
rearrange the amplitude in the cross-channel, as depicted in
Fig.1(b). Since channel duality is the feature observed in
string theory, as in the noncommutative field theory computations,
we express the Feynman diagram in terms of Schwinger time
variables by expressing the $i$-th propagator in Eq.(\ref{npt}) as
\bean \int \d s_i \exp \left[ - s_i \O m \Delta \Big(p + \sum_{j=1}^i
k_j, \pi + \sum_{j=1}^i \kappa_j\Big) \right]. \label{propagator}
\eean
The Schwinger parameter $s_i$ can be thought of as the length of the
$i$-th arc,  or the distance between the $(i-1)$-th and $i$-th vertices in
the Feynman graph. For the $N$-point amplitude, we have $N$
propagators and hence $N$ Schwinger parameters. These Schwinger
parameters span the moduli space of $N$ marked points on the
one-loop vacuum diagram. Equivalently, 
%using Killing symmetry on a circle, 
they can be viewed as parametrizing the moduli space of the 
perimeter of the one-loop vacuum diagram and $(N-1)$ relative
marked points on a circle of unit radius. The perimeter is
measured by $s \equiv (s_1 + \cdots + s_N)$.

With Schwinger parameters introduced, integration over the loop
momentum $P$ is reduced to a Gaussian integral. In doing so, as seen
from the product of $N$ propagators Eq.(\ref{propagator}), the
coefficient of $P^2$ is $s$. So, factoring out $s$ from the
complete-square for $P$, the residual terms are proportional to
$\sigma_i \equiv (s_i/s)$. These $\sigma_i$'s are precisely the
moduli of relative marked points on a unit circle. Integration
over $P$ now yields $s^{-2} \times s$ from bosonic and fermionic
integrals.

The residual terms in the exponent depends on the external momenta
$K_i$ in a complicated and unilluminating way. We are
however interested in the effective superpotential -- terms that
do not depend on derivatives other than those in the
$\star$-product. So we will isolate these terms by taking the
limit of large non(anti)commutativity and small external momenta:
\bean k_i, \kappa_i \rightarrow {\cal O}(\epsilon) \qquad {\rm and}
\qquad \Theta^{\mu \nu}, \, C^{\alpha\beta} \rightarrow {\cal
O}(\epsilon^{-2}). \label{ourlimit}\eean
In fact, we shall see later that this limit is quite harmless,
because we will still reproduce the known results in the limit
that either $\Theta$ or $C$ is taken back to zero (along with
$\lambda$, since this feature is new in our analysis). The result
for the $N$-point Green function is quite simply
\bea\label{greenf} G_{N} & = & {1 \over 4 \pi^2 } {\O m \over
4 \lambda} {( -g \O m)^N \over N!} \int \prod_{i=1}^N \d^6 K_i \,
\widetilde{\Phi}_0(K_i) \, \delta^6(K_1 + \cdots + K_N) \nonumber\\
&\times& \int_0^\infty \d s ~s^{N-2} \exp \Big[-s \, m \O m- {1
\over 4 s} 
\Big((\Theta k_0)^2  - {4\lambda \over \O m}
(C \kappa_0)^2 \Big) \Big]\\
&\times& \int_0^{\sigma_1} \d \sigma_2 \cdots \int_0^{\sigma_{N-1}}
 \d \sigma_N
\left( e^{+ i(\Theta k_0) \sum_k \sigma_k k_k + {i \over 2}\sum_{k
\leq j}{\epsilon_j + \epsilon_k \over 2}k_j \wedge k_k}\right)
\left( e^{ - (C \kappa_0) \sum_k \sigma_k \kappa_k + {1 \over 2}\sum_{k
\leq j}{\epsilon_j + \epsilon_k \over 2}\kappa_j \wedge \kappa_k}\right).
\nonumber \eea
We identify two separate pieces: the vacuum moduli
contribution in the second line, and the phase factors from
non(anti)commutativity in the third line.
%Contributions from twisted and untwisted vertices can be grouped
%separately and have opposite signs of
%${\epsilon_j + \epsilon_k \over 2}$.
%
Appendix B contains some details of the derivation of
Eq.(\ref{greenf}).

In Eq.(\ref{greenf}) above, the integration over the relative
moduli is specified for a particular ordering of given $n$
untwisted and $(N-n)$ twisted background field insertions. If we
sum over all possible orderings, the effect is just to extend
every $\sigma_i$ integral over $[0, 1]$. Moreover, and
remarkably, the phase factor in Eq.(\ref{greenf}) is factorized
into those for untwisted insertions and those for twisted ones.
This is very much like open string annulus amplitudes, and
eventually enables us to re-express the amplitudes in
the cross-channel as in Fig.1(b).

The effective superpotential is obtained by summing all $N$-point
contributions. Taking into account the combinatoric factor ${N
\choose n}$ for the $N$-point Green function with $n$ untwisted vertices, 
the result is
%Including a factor of ${N \choose n}$ for the number of ways to choose $n$ twisted vertices,
%
\bea\label{pathintegral} W_{\rm eff}[\Phi_0] &= & {m \O m\over 4
\pi^2} \int_0^{\infty} {\d \tau \over \tau^{2}} \int \d^6 K_0
{\cal W}_{K_0,\tau}[\Phi]  {\cal K}(K_0;\tau) {\cal
W}_{-K_0,\tau}[\Phi], \eea
where we have introduced the dimensionless overall modulus $\tau=m \O
m s$; the `scalar' open Wilson line
%\footnote{This expression is
%for an untwisted open Wilson line. For a twisted open Wilson line
%we change the sign of the phase factors.}
%
\bea \label{S-line} {\cal W}_{K_0,\tau}[\Phi]=\Tr \int
{\d^6 Y}{\cal P}_\sigma  
\exp_\star \left(  - {g \tau\over m} \int_0^1 \d \sigma
~\Phi(y+(\Theta k_0)\sigma, ~\theta-(C \kappa_0)\sigma) \right)\star
e^{-i k_0 \cdot y-\kappa_0\cdot\theta}, \eea
where `Tr' refers to trace over color indices in case the scalar
superfield $\Phi$ is taken to be $U(N)$ matrix-valued; and the
`cross-channel propagator'
\bea \label{kernel}
 {\cal K}(K_0; \tau)=
\left({\O m \over 4 \lambda} +{L\over \tau} \kappa_0^2 \right) \exp
\left( - \tau - {M^2 \over \tau} \right), \eea
where $L, M$ are dimensionless combinations of the
non(anti)commutativity parameters:
\bean  L \equiv {1 \over 4} m \O m \det(C), \qquad M \equiv {1 \over 2}
 {\sqrt{m \O m} | \Theta  k_0|}.
\eean
It is worth noting that the open Wilson line defined by
Eq.(\ref{S-line}) depends on the overall modulus $\tau$, and the
final result is the weighted sum of all open Wilson lines in
$\tau$-space. Roughly speaking, the parameter $\sigma$ is the
affine parameter around each index-loop, and the parameter $\tau$
sets the size of the index-loop in superspace.

The manifestly channel-dual expression Eq.(\ref{pathintegral}) of
the one-loop effective superpotential is the central result. As in the
noncommutative field theories, by identifying the open Wilson line
as a closed string field, the effective action takes strikingly
the same form as the quadratic term in the closed string field theory.
In fact, adopting the result \cite{leeetal}, we expect that the
$\ell$-loop contribution to the effective superpotential is
expressible as an interaction involving $\ell$ closed string
fields.

 We will base our discussion on
Eq.(\ref{pathintegral}) and explore further aspects of the channel
duality.

%%%%%%%%%%%%%%%%%%%
\section{UV-IR mixing}
%%%%%%%%%%%%%%%%%%%%%
Having obtained the effective superpotential, in this section, we
shall discuss how physics changes in various limits of coupling
parameters. The first case is that both $\Theta$ and $\lambda$ are
nonzero. In this case, we can integrate the overall modulus $\tau$
and simplify the effective superpotential Eq.(\ref{pathintegral})
further. The second case is that $\Theta=0$ (so the $y$
coordinates commute).  Here we observe the dipole effect in superspace.
The third case is that
$\Theta=0$ and $\lambda \rightarrow \infty$, where the open Wilson line collapses to a point.  We recover the effective superpotential of the Wess-Zumino model deformed by $C$ only.

%%%%%%%%%%%%%%%%%%%%%%%
\subsection{Full non(anti)commutativity}
%%%%%%%%%%%%%%%%%%%%%%

Consider first the situation when we have both noncommutativity of
Grassmann-even coordinates and nonanticommutativity of Grassmann-odd
coordinates. We can then perform integration over the overall
modulus $\tau$ in Eq.(\ref{greenf}) or, equivalently, in
Eq.(\ref{pathintegral}) after expanding the open Wilson line
${\cal W}[\Phi]$ in powers of $\Phi$'s and bringing down $\tau$'s
from the exponent. The integration involved is of the form
\bea \label{integral-S}
\int \d\tau \tau^{N-2}  {\cal K}(\tau,K_0)=
2 M^{N-1} \left[{\O m \over 4 \lambda}{\bf K}_{1-N}(2M)
+ \kappa_0^2 {L \over M}
{\bf K}_{2-N}(2M) \right],
\eea
where ${\bf K}_n(z)$ is the modified Bessel function. A relevant
property of the modified Bessel function is that asymptotically at
large $z$,
\bean
{\bf K}_n(z)\sim \sqrt{\pi \over 2 z} e^{-|z|}[1+{\cal O}({1\over |z|}),
\eean
which is {\it independent} of $n$. So,
in the limit we are taking Eq.(\ref{ourlimit}), integrating
explicitly over the overall modulus results in
\bea \label{Gamma-S-int}
\int \d^6 Y \, W_{\rm eff} = \left({m \O
m \over 4 \pi^2}\right)
\int \d^6 K_0 W_{K_0}[\Phi]_U
\left({\O m \over 4 \lambda M} +
{L \over M^2}\kappa_0^2
\right)
{\bf K}_{0}(2M)  W_{-K_0}[\Phi]_T
\eea
where we have defined a new open Wilson line with superspace size
determined by $M$, via the kinematic factor $\sqrt{h} \equiv {g \over m}M$:
\bea\label{new-Wilson}
 W_{K_0}[\Phi]_U={\cal P}_t\left[-
\int  \d^6 Z
\exp\left( \int_0^1 \d\sigma
 \sqrt{h} \Phi(y+\sigma(\Theta k_0),\theta-\sigma(C\kappa_0))
\right) \star e^{-i k_0 \cdot y - \kappa_0 \theta }\right]_U,
\eea
and the twisted function $W_{K_0}[\Phi]_T$ just has an extra
minus sign in every phase factor.

It is instructive to compare the general expression of the
effective superpotential Eqs.(\ref{pathintegral}, \ref{S-line})
with the present one, Eqs.(\ref{Gamma-S-int}, \ref{new-Wilson}).
 In fact, the expression
Eq.(\ref{new-Wilson}) is almost identical with the one given in
\cite{wilsonline} for ordinary noncommutative field theory, where
$\sqrt{h}={g  M\over m} $ 
plays the role of the metric factor in one-dimensional
parameter space and equals $|\dot y(\sigma)|$. In the latter,
$y(\sigma)$ sets the metric factor since it is a translation-invariant
interval in noncommutative space, so we should expect the
appearance of a supertranslation-invariant interval in the present
case. Indeed it is so and is determined by $M$, as we now explain.

In the ordinary ${\cal N}=1$ superspace $(x,\theta,\O \theta)$,
the supertranslation invariant interval is defined by
\bea (\Delta s)^2 =  w^\alpha  w_\alpha \qquad \mbox{where} \qquad
w =(x_2-x_1) + i\theta_1 \sigma (\O \theta_2-\O \theta_1) -i(
\theta_2- \theta_1) \sigma \O \theta_1. \nonumber \eea
It is invariant under a general supertranslation,
\bea x^m \rightarrow x^m + i(\theta\sigma \O \zeta-\zeta \sigma \O
\theta)^m + \varepsilon^m,~~~~ \theta\rightarrow
\theta+\zeta,~~~\O \theta\rightarrow \O \theta+\O \zeta. \nonumber
\eea
However, once the ${\cal N}=1$ superspace is reduced to ${\cal
N}=1/2$ by the non(anti)commutativity, the latter's chiral
coordinates $(y=x+i\theta\sigma \O \theta,\theta)$ are the unique
choice of local coordinates. This is because the reduced
supersymmetry then requires that $\O \zeta=0$, and $\O\theta$ is
no longer a part of ${\cal N}=1/2$ superspace coordinates. It then
follows that $\Delta y$ is the unique interval invariant under
supertranslation in ${\cal N}=1/2$ superspace:
\bea y^m \rightarrow y^m + \varepsilon^m, \qquad \theta^\a
\rightarrow \th^a + \zeta^\a. \nonumber \eea

Now that we have understood that $\Delta y$ is the only ${\cal
N}=1/2$ supertranslation invariant length, emergence of the
characteristic length $\dot y(\sigma)= |\Theta k_0|$ in the
open Wilson line Eq.(\ref{new-Wilson}) is readily understood.
 Recall the modulus integral
\bean
\int \d\tau \tau^{N-2} e^{-\tau-{M^2\over \tau}}.
\eean
Because both $\tau$ and ${1\over \tau}$ appear in the exponential, the
dominant contribution comes from the saddle point, where
\bean
\tau={M^2\over \tau} \Longrightarrow \tau=M
\eean

%%%%%%%%%%%%%%%%%%%%%
\subsection{Nonanticommutative Limit}
%%%%%%%%%%%%%%%%%%%%%
We next consider the limit $\Theta^{mn} = 0$.
In this limit, only the $\th$ coordinate is nonanticommutative while
the $y$ coordinate is commutative. In this case, we should {\it not}
do the overall moduli integration $\int \d \tau$, but keep
the form (\ref{pathintegral})
(\ref{S-line}). The reason is that since $M=0$,
there
is no ${1\over \tau}$ term in the exponential. Thus there is {\it no}
characteristic length to dominate the integral, and we must
sum all contributions over $\tau$. This explains also why $\tau$ shows
up in the definition of the open Wilson line.
The dependence of the open Wilson line on the overall modulus $\tau$
is new compared to the familiar understanding from
the ordinary noncommutative field theory \cite{owl}.

Besides the new form of the open Wilson line, there are also a few
new features related to the fermionic non-anticommutativity. Like the
dipole effect in the ordinary noncommutative field theory
\cite{dipole}, we see here the dipole effect in $\theta$
coordinates too.  This is quite generic.
Consider, for instance, a `super-dipole' whose dipole moment is
proportional to $\zeta$. Denote the dipole's constituents by a wave
function $\Phi(\th)$. Then, the form-factor of the `super-dipole' is
given by Fourier transform:
\bean
I& \equiv & \int \d^2 \theta \Phi(x,\theta) \star
\Phi(x,\theta+\zeta) \star e^{\kappa_0 \theta}\\
& = & (-4) \int  \d^2 \kappa_2 \Phi(x,\kappa_0-\kappa_2) \Phi(x,\kappa_2)
e^{-\kappa_2 \zeta} e^{-{1\over 2} \kappa_0\wedge \kappa_2}.
\eean
If the distribution is $\Phi(x,\theta)=\Phi(x) \delta^2(\theta)$,
we have $ \Phi(x,\kappa)=\Phi(x)$, so that
\bean
I & = & \Phi(x)^2 \delta^2(\zeta+ {1\over 2}(C\kappa_0))
\eean
Thus, the dipole moment is not fixed but is proportional to the
center-of-mass momentum $\kappa_0$:
\bean \Delta \th^\a = \zeta^\alpha \sim C^{\alpha\beta}
\kappa_{0\beta}, \eean
exhibiting precisely the dipole relation Eq.(\ref{dipole}).

%%%%%%%%%%%%%%%%%%%%%%%%
\subsection{Deformed supersymmetry limit}
%%%%%%%%%%%%%%%%%%%%%%%%
It is also of interest to take the limit $\lambda \rightarrow
\infty$. This is the limit that the ${\cal N}= 1/2$ kinetic
supermultiplet is suppressed and that the ${\cal N}=1$
supersymmetry is
restored through smooth extrapolation
as the
non(anti)commutativity $C^{\a\b}$ is taken to zero. For simplicity, 
we will also take $\Theta^{mn} \rightarrow 0$.

In the $\lambda \rightarrow \infty$ limit, the concerned part of the
cross-channel propagator is
\bea  \label{lambda=0}  
\left[ {\O m\over 4 \lambda}+
{L \over \tau} \kappa_0^2
\right]
 \label{parts}
\eea
in Eq.(\ref{pathintegral}). In terms of superspace coordinates,
$\kappa_\a = Q_a$, so in the propagator, there is one term with an
insertion of $Q^2$, and one without. We know \cite{bfr} that the
non(anti)commutative $\star$-products produce only even powers of
$Q^2$ (because $C^{\a\b}Q_{\a}\Phi Q_{\b}\Phi=0$). With both of
these terms present, all possible powers of $Q^2$ are produced,
both odd and even (up to one less than the total number of
$\Phi$'s). Thus, all terms generated in $W_{\rm eff}$ appear
already at one loop.

In the $\lambda \rightarrow \infty$ limit,  the second term
in Eq.(\ref{parts}) dominates, so the $\Phi Q^2 \Phi$ term
disappears, and the theory reduces to the classical part of the
deformed Wess-Zumino model discussed in \cite{seiberg}.  One loop
corrections give only terms like $\Phi^k (Q^2\Phi)^{2l+1}$, as
observed in \cite{bfr,ty}. Also, since $\kappa_0^2$ appears in the
cross-channel propagator explicitly, we can not bring any further
factors of $\kappa_0$ from the open Wilson line. Thus, one can set
the Grassmann coordinate $(\theta - C \kappa_0
\, \sigma)$ of the open Wilson line 
effectively to $\theta$, viz. the open Wilson line is
collapsed to a local operator (with a base point at $\theta$).
This explains why we do not see the open Wilson line if the
kinetic multiplet term is absent. If in addition $\Theta=0$, the
effective superpotential reads
\bea \int \d^6 Y \, W_{\rm eff}[\Phi_0] = {(m \O m)^2 \over 4
\pi^2} \int_0^\infty {\d \tau \over \tau^3} e^{-\tau} \int \d^6 Y
\, {\cal W}_\tau[\Phi_0] \Big(-{1 \over 4} Q^2 \Big) {\cal
W}_\tau[\Phi_0] \label{0eff} \eea
where
\bea {\cal W}_\tau[\Phi(Y)] = {\rm Tr} \exp_\star \left[- \tau {g
\over m} \Phi_0(Y) \right]. \nonumber \eea
is the open Wilson line, now collapsed to a point.

As it stands, the overall-modulus integral in Eq.(\ref{0eff}) is
divergent at $\tau \sim 0$ for terms up to quadratic in $\Phi_0$.
This can be understood from the observation that, as $\Th^{mn}$ is
now set to zero, nonplanar diagram of Grassmann-odd $\star$-product
no longer exhibits the UV-IR mixing for Grassmann-even momenta. As
the divergence would be absent in case $\Th^{mn}$ is nonzero, a
consistent treatment would be to treat the logarithmically
divergent operator $\Phi Q^2 \Phi$ as a state created by the open
Wilson line on the mass-shell. In other words, with a change of
variables $s = 1/\tau$, the logarithmic divergence originates from
$\int \d s \exp(-P^2 s)$ at $P^2 \sim 0, \,\, s \sim \infty$.

%%%%%%%%%%%%%%%%%%%%%%%

%%%%%%%%%%%%%%%%%%%%%%%%
\section{Planar Contribution}
%%%%%%%%%%%%%%%%%%%%%%%%
So far, we have focused mainly on the nonplanar contribution to the
effective superpotential $W_{\rm eff}$. For example, the $N$-point
function involves $n$ untwisted U and $(N-n)$ twisted T
interactions, attached in double-line notation to inner and outer
index loops, respectively. The planar contributions originate from
two exceptional cases, $n=0$ and $n=N$. In the double-line
notation, they correspond to either all U or all T interactions,
attached only one of the two index loops. Equivalently, they
correspond to setting either of the two open Wilson lines in the effective superpotential
Eq.(\ref{pathintegral}) to
unity. Since the corresponding index loop is free, these planar
amplitudes would be proportional to Tr $ 1 = N_c$.

\subsection{Planar combinatorics}
Consider again the combinatorics. Because the free
index-loop amounts to setting the corresponding open Wilson line
to unity, we find for the net-momentum flow $k_0=0$ and $\kappa_0 = 0$. This vanishing has
two effects. First, the path-ordered $\star$-product reduces
to the standard $\star$-product Eq.(\ref{starproduct}). Second,
the argument of the modified Bessel function in the cross-channel
propagator vanishes.  More precisely, if we introduce an ultraviolet cutoff
$\Lambda$, then the argument needs to be set to $\Lambda^{-2}$, and
in the end we take the limit $\Lambda \rightarrow \infty$. First we
recall the small argument behavior of the modified Bessel function:
\bea {\bf K}_n (2M) \sim {1 \over 2} \Gamma(|n|) M^{-|n|}
\label{small1} \eea
for $n \ne 0$, and
\bea {\bf K}_0 (2M) \sim - \log \, M \label{small2} \eea
for $n=0$. One again finds ultraviolet divergence for lower-point
amplitudes, so we implicitly perform a suitable renormalization.

The small argument behavior Eqs.(\ref{small1}, \ref{small2})
concerning the cross-channel propagator can be understood directly
from the overall modulus integral. Expansion of the open Wilson
line contributes to the $N$-point function a factor of $\tau^N/N!$, so
the overall-modulus integral yields
\bea \int {\d \tau \over \tau^2} \, {\tau^N \over N!} \sim {(N-2)!
\over N!} \nonumber \eea
We then decompose this factor into partial fractions,
\bea {(N-2)! \over N!} = {1 \over N(N-1)} = -({1 \over N} - {1
\over (N-1)}), \nonumber \eea
and resum each term separately over $N$. Thus, the planar contribution
to the effective superpotential is now given by
\bea W_{\rm eff}[\Phi_0] \Big|_{\rm planar} \simeq  - N_c \, (m +
g \Phi_0) \star \log_\star (m + g \Phi_0). \label{planar}
\eea

The functional form of the planar contribution Eq.(\ref{planar}) is
interesting. Suppose we had started with, instead of the action
Eq.(\ref{holoWZ}), a new action in which $Q_\a$ is replaced by $Q^2
+ W^\a W_\a$, where $W_\a$ refers to a spinorial background field.
The background field may be viewed as a gauge field associated
with super-translation. We then observe diagrammatically that the
net effect is to replace $g \Phi_0 \rightarrow g \Phi_0 + {\cal
W}^\a {\cal W}_\a$. We can then even turn off the background field
$\Phi_0$, set $m=0$,  and obtain the planar contribution to the effective
superpotential of ${\cal W}^\a {\cal W}_\a$ as
\bea W_{\rm eff}[{\cal W}^\a {\cal W}_\a] \Big|_{\rm planar} \sim
- N_c \, {\cal W}^a {\cal W}_\a \log \Big({\cal W}^\a {\cal W}_\a
\Big). \nonumber \eea
Intriguingly, this is precisely the same functional form as
the Veneziano-Yankielowicz glueball superpotential \cite{vene}, 
if we identify $W_\a$ as the gauge superfield. This
may be an indication that even the non-analytic part of the
nonperturbative glueball superpotential may be derived from
perturbative dynamics in the context of the Dijkgraaf-Vafa
correspondence. This is not inconceivable, since the logarithmic
term in Eq.(\ref{planar}) originates essentially from dimensional
transmutation, viz. the Coleman-Weinberg mechanism, and this is what
also underlies the original derivation of the glueball
superpotential \cite{vene}.

%%%%%%%%%%%%%%%%%%%%%%%%%%%%%%%%%%%%%%%%%%%%%%%%%%%%%%%%%%%%%%%%%%%%%
\section*{Acknowledgements}
We are grateful to N. Seiberg for enlightening discussions. R.B.
and B.F. were supported by the NSF grant PHY-0070928.  R.B. thanks Harvard 
University for hospitality during the final stage of this work.  
S.J.R. was
supported in part by the KOSEF Interdisciplinary Grant
98-07-02-07-01-5 and the KOSEF Leading Scientist Grant.  S.J.R.
was a Member at the Institute for Advanced Study during this work.
He thanks the School of Natural Sciences for hospitality and for
the grant in aid from the Fund for Natural Sciences.

%%%%%%%%%%%%%%%%%%%%%%%%%
\appendix
\section*{Appendix}
\section{Notations and conventions}
%%%%%%%%%%%%%%%%%%%%%%%%
\subsection{Fourier transform in superspace}
Here, we collect our notation and present some useful formulas.
The ${\cal N}=1/2$ superspace coordinates and momenta are
abbreviated as $Y^a$ and $K_a$, respectively
\bea Y^a \equiv (y^m,\theta^\a) \qquad \mbox{and} \qquad K_a
\equiv (ik_m,\kappa_\a). \nonumber \eea
We adopt the following convention of the superspace Fourier
transformation:
\bea \widetilde{\Phi}(K) & = & \int \d^4 y ~\d^2 \theta \,\, e^{-i
k y - \kappa \th} \, ~\Phi(Y) \,
\equiv \int \d^6 Y e^{ - K Y} ~\Phi(Y) \nonumber \\
\Phi(Y) &= &\int \!\! {\d^4 k\over (2\pi)^4} {\d^2\kappa \over
(i/2)^{2}} e^{ik y+\kappa\theta} \widetilde{\Phi}(K)\equiv \int
\d^6 K ~e^{KY} ~\widetilde{\Phi}(K). \nonumber \eea
In the convention adopted, we abbreviate the Dirac
$\delta$-functions as
\bea \int \d^6 K  ~e^{K Y} & = & \delta^4(y)
\delta^2(\theta) \equiv \delta^6(Y ) \nonumber \\
\int \d^6 Y ~e^{-KY} & = & (2\pi)^4 \delta^4(k) (i/2)^2
\delta^2(\kappa) \equiv \delta^6(K). \nonumber \eea

%%%%%%%%%%%%
\subsection{The Lagrangian in momentum space}
%%%%%%%%%%%%
The Lagrangian is given by
\bea S=\int \d^6 Y {1\over 2} \Phi(Y) \left[{1 \over 4\lambda}
Q^2-{1 \over \O m}\Box + m \right] \Phi(Y)+{g\over 3} \Phi(Y)
\star \Phi(Y) \star\Phi(Y). \nonumber \eea
The propagator is given by
\bea \langle \Phi(Y)\Phi(Y')\rangle & = & \Big[ -{1 \over \O m}
\Box_y + {1 \over 4 \lambda} Q^2 +m \Big]^{-1} \delta^6(Y-Y')
\nonumber \eea
in configuration superspace, or equivalently in momentum space,
 \bea \langle \widetilde{\Phi}(K)\widetilde{\Phi}(K')\rangle & = &
 \Big[{1 \over \O m} k^2 +{1 \over 4\lambda} \kappa^2  + m \Big]^{-1}
 \delta^6(K-K'). \nonumber
\eea
Using these results, the Lagrangian in momentum superspace is
obtained as
\bea S & = & {1\over 2} \int \d^6 K_1 \d^6 K_2 \delta^6(K_1+K_2)
\Big[{1\over \O m} k_1^2 + {1 \over 4\lambda} \kappa_1^2 + m \Big]
\widetilde{\Phi}(K_1) \widetilde{\Phi}(K_2) \nonumber \\
&+& {g\over 3} \int \prod_{i=1,2,3}\d^6 K_i \widetilde{\Phi}(K_i)
~\delta^6(K_1+K_2 + K_3) e^{-{1\over 2}\sum_{i<j} \kappa_i\wedge
\kappa_j-{i\over 2}\sum_{i<j} k_i\wedge k_j}, \nonumber \eea
where we have defined
\bea \kappa_1\wedge \kappa_2 & = & \kappa_{1\alpha} C^{\alpha
\beta} \kappa_{2\beta} =-\kappa_2\wedge \kappa_1,
~~~~~~~(C\kappa)^\alpha =  C^{\alpha \beta}\kappa_{\beta},
\nonumber \\
p_1\wedge p_2 & = & p_{1m} \Theta^{mn} p_{2n}= -p_2\wedge p_1,
~~~~~~~~(\Theta p)^m=\Theta^{mn} p_{n} \nonumber \eea
In fact we can define
\bea K_1 \wedge K_2 \equiv (ik_1)_m  \Theta^{mn}(ik_2)_n
+\kappa_{1\alpha} iC^{\alpha \beta} \kappa_{2\beta}, \nonumber
\eea
and then the phase factor is just $\sum_{i<j} {i\over 2} K_i
\wedge K_j$. However, since we may deal with the bosonic and
fermionic parts separately, we will use the component form or
compact form as convenient.

%%%%%%%%%%%%%%%%%%%%
\section{Evaluation of the momentum integrals}
%%%%%%%%%%%%%%%%%%%%

Here we present some details of the derivation of
Eq.(\ref{greenf}).

We have defined the net-flow momenta across the channel
\bean k_0 \equiv \sum_{i \in \{U\}} k_i = -\sum_{i \in \{T\}} k_i,
\qquad \kappa_0 \equiv \sum_{i \in \{U\}} = - \sum_{i \in \{T\}}
\kappa_i, \eean
where $i \in \{U\}/\{T\}$ means that vertex $i$ is un/twisted.
Then,
\bean \sum_{i=1}^N {\epsilon_i \over 2} k_i = k_0, \qquad
\sum_{i=1}^N {\epsilon_i \over 2} \kappa_i = \kappa_0. \eean
Using the antisymmetry of the wedge product and the overall
momentum conservation, one can see that
\bean \sum_{i=1}^N {i \epsilon_i \over 2}  k_i \wedge(p
+\sum_{j=1}^{i} k_j) =i k_0 \wedge p + {i \over 2}\sum_{j \leq
i}{\epsilon_i + \epsilon_j \over 2} k_i \wedge k_j \eean
and similarly
\bean \sum_{i=1}^N { \epsilon_i \over 2} \kappa_i \wedge(\pi
+\sum_{j=1}^{i} \kappa_j) = \kappa_0 \wedge \pi + {1 \over
2}\sum_{j \leq i}{\epsilon_i + \epsilon_j \over 2}\kappa_i \wedge
\kappa_j \eean
Move the propagator into the exponent by introducing Schwinger
parameters $s_j$:
\bean G_{N} & = &  -{1\over N!} (-g \O m)^N \int \prod_{i=1}^N
\d^4 k_i \d^2 \kappa_i \widetilde{\Phi}_0(k_i,\kappa_i)
\delta^4(k_1+\cdots+k_N) \delta^2(\kappa_1+\cdots+\kappa_N)\\
&\times& \int \d^4 p  \d^2 \pi \int_0^\infty \d s_1 \cdots \d s_N
\prod_{i=1}^N \exp\Big[-s_i \Big(  (p+\sum_{j\le i}
k_j)^2 +{\O m \over 4\lambda} (\pi +\sum_{j\le i} \kappa_j)^2 + m\O m
\Big)\Big]\\
&\times& \prod_{i=1}^N \exp \left[ i k_0 \wedge p + {i \over
2}\sum_{j \leq i}{\epsilon_i + \epsilon_j \over 2}k_i \wedge k_j
+\kappa_0 \wedge \pi + {1 \over 2}\sum_{j \leq i}{\epsilon_i +
\epsilon_j \over 2}\kappa_i \wedge \kappa_j \right]. \eean
To continue, we group the phase factors into bosonic and fermionic
parts.
\begin{itemize}
\item (a). The phase factor of the bosonic part  is given by
\bean & & -p^2(\sum_i s_i)- 2 p\sum_i s_i \sum_j k_j -\sum_i s_i
(\sum_j k_j)^2 + i k_0 \wedge p + {i \over 2}\sum_{j \leq
i}{\epsilon_i + \epsilon_j \over 2}k_i \wedge k_j.
 \eean
Complete the square and perform the loop momentum $p$-integral.
This gives an overall factor of ${\Omega_3 / 2(\sum_i s_i)^2}$
where $\Omega_3$ is the volume of unit 3-sphere. Redefine the
moduli parameters as follows:
\bea s \equiv \sum_{i=1}^N s_i; \qquad \sigma_i \equiv {1 \over s}
\sum_{i=i}^N s_i. \nonumber \eea
In these new variables, we have
\bean \sum_i s_i \sum_j k_j & = & s \sum_i \sigma_i k_i\\
\sum_i s_i (\sum_j k_j)^2 & = & s \sum_{i=1}^N \sigma_i k_i
(-k_i+ 2\sum_{j=1}^i k_j) \\
\int_0^\infty \d s_1 \cdots \d s_N &=& \int_0^\infty \d s~s^{N-1}
\prod_{i=2}^N \int_0^{\sigma_{i-1}} \d \sigma_i \eean
The remaining bosonic phase can be rewritten as
\bean s (\sum_i \sigma_i k_i)^2 - s \sum_i \sigma_i k_i (-k_i +
\sum_{j \leq i} k_j) + i(\Theta k_0) \sum_i \sigma_i k_i - {1
\over 4 s}(\Theta k_0)^2 + {i \over 2}\sum_{j \leq i}{\epsilon_i +
\epsilon_j \over 2} k_i \wedge k_j.\eean

\item (b). The phase factor of the fermionic part is given by
\bean & & - {\O m \over 4\lambda} (\sum s_i) (\pi + \sum_j
\kappa_j)^2 + \kappa_0 \wedge \pi + {1 \over 2}\sum_{j \leq
i}{\epsilon_i + \epsilon_j \over 2}\kappa_i \wedge \kappa_j. \eean
The result of the $\pi$ integral, in terms of the new moduli, is
an overall factor of $- (\O m / 4\lambda) s$ and a phase of
\bean s{\O m \over 4\lambda} (\sum_i \sigma_i \kappa_i)^2 +2 s {\O
m \over 4\lambda} \sum_i \sigma_i \kappa_i \sum_j \kappa_j - (C
\kappa_0) \sum_i \sigma_i \kappa_i + {4\lambda \over 4 \O m} {(C
\kappa_0)^2 \over s} + {1 \over 2}\sum_{j \leq i}{\epsilon_i +
\epsilon_j \over 2}\kappa_i \wedge \kappa_j. \eean

\end{itemize}

\noindent Now we take the limit of large non(anti)commutativity
and small external momenta:
\bean k_i, \kappa_i \rightarrow {\cal O}(\epsilon) \qquad
\mbox{and} \qquad \Theta^{mn}, C^{\alpha\beta} \rightarrow {\cal
O}(\epsilon^{-2}). \eean
Evidently only the last three terms in each case  remain. Equation
Eq.(\ref{greenf}) thus follows immediately.

%%%%%%%%%%%%%%%%%%%%%%%%%%%%%%%%%%%%%%%%
%========
%%%%%%%%%%%%%%%%%%%%%%%%%%%%%%%%%%%%%%%%

\bibliographystyle{JHEP}

\end{document}